\documentclass[
reprint,
 amsmath,amssymb,
 aps,
pra,
]{revtex4-1}

\usepackage{graphicx}
\usepackage{dcolumn}
\usepackage{bm}

\usepackage{braket} 
\usepackage{mathrsfs} 
\usepackage{physics} 
\usepackage[retainorgcmds]{IEEEtrantools} 
\usepackage[caption=false]{sub fig} 
\usepackage{lipsum} 
\usepackage{amsthm} 
\usepackage{dsfont} 
\usepackage{accents} 

\newcommand{\eqnref}[1]{eq.~\ref{#1}}

\newcommand{\figref}[1]{Fig.~\ref{#1}}

\newcommand{\ubar}[1]{\ensuremath{\underaccent{\bar}{#1}}}

\newlength{\argl}
\newlength{\argh}
\newlength{\tildel}
\newlength{\tildec}

\begin{document}

\preprint{APS/123-QED}

\title{Quantum Mechanics in Technicolor; Analytic Expressions for a Spin-Half Particle Driven by Polychromatic Light}

\author{B. Yuen}
 \email{benjamin.yuen@physics.ox.ac.uk}
\affiliation{%
Clarendon Laboratory\\
Parks rd\\
Oxford\\
OX1 3PU
}%


%

\date{\today}

\begin{abstract}
A vast collection of light-matter interactions are described by the single-frequency Rabi model. However, the physical world is polychromatic, and until now there is no general method to find analytic solutions to the multi-frequency Rabi model. We present the first general method to solve the Rabi model with $N$ frequency modes.
Analytic expressions are derived for a spin-half particle in a multi-frequency coherent field, and accurately describe the time evolution even when the interactions are strong.
The dynamics are solved in an extended dressed atom picture whereby the spin is progressively dressed by each frequency mode.
For weak fields, the closed form expressions for the time evolution of atoms or their analogues take a particularly simple form. 
These results analytically solve problems in a wide range of fields including quantum information processing, coherent control and resonant nonlinear optics.

\end{abstract}

\pacs{Valid PACS appear here}
\maketitle

In 1937 Rabi introduced a model for a two-level atom driven by a monochromatic classical field \cite{Rabi1937}, now ubiquitous in physics. Despite its simple form the second quantised Rabi model has only been shown to be integrable within the last decade \cite{Braak2011}. Solutions for linear \cite{Braak2011, Chen2012, Zhong2013} and anisotrophic polarised \cite{Xie2014} quantum driving fields have followed, and simple closed form approximate solutions are still emerging \cite{Xie2017}.
Famously, Jaynes and Cummings \cite{Jaynes1963} were first to consider the quantum model in 1963 and its elegant solutions under the rotating wave approximation. These exhibit many interesting phenomena from collapses and revivals to preparation of Schr\"odinger cat states \cite{Shore1993, Haroche2013}. Soon after Cohen Tannoudji and Haroche formalised the dressed atom picture to give a comprehensive description of atom-photon interactions \cite{Cohen66, Cohen69, Cohen92}. 

The physical world is rich with colour and appears monochromatic only in isolated situations. It is therefore necessary to develop a general, non-perturbative framework for polychromatic atom-photon interactions. A two mode, two photon variant of the Jaynes Cummings model was considered in\cite{Gou1989}, and shown to produced inter-modal correlations, squeezing and population revivals \cite{Shore1993}. Polychromatic interactions have been studied in other specific cases (e.g. \cite{Metcalf2017}), and several analytic solutions have been proposed to the optical Bloch equations with periodic classical driving fields \cite{Barnes2012, Wu2007, Barata2000}. 

The broadest collection of known polychromatic phenomena is in non-linear optics \cite{Boyd2003}. Armstrong et al. predicted an impressive range of monochromatic and polychromatic non-linear optical effects with a perturbative semi-classical theory \cite{Armstrong1962}. However, a different approach is needed for resonant or strongly coupled systems where perturbation theory diverges.

The extension of the Rabi model to polychromatic fields is motivated by wealth of phenomena discovered for the analogous generalisation from two level atoms to multiple levels. A polychromatic theory could immediately be applied to pulse shaping which is used to enhance coherent control and measurement in fields from NMR to atom interferometry, and currently relies on numerical optimisation. A general non-perturbative approach to solving the polychromatic Rabi model would describe strongly coupled and near resonant non-linear optical systems.
Quantum information processing would benefit from polychromatic driving fields which have sufficient degrees of freedom for fast targeted control of qubits within an array without the need to strongly lift their degeneracy \cite{Xia2015,Wang2015}.


This letter presents a general method to find analytic solutions to the polychromatic Rabi
\begin{equation}
	H= 
		\tfrac12 \omega_0 \sigma_z  + \sum_{k} \left[a_k^{\dagger} a_k k \omega_f
		+g \left(u_k a_k + u_k^* a_k^{\dagger} \right) \sigma_x \right]
\end{equation}
where $\pm \omega_0$ are the spin-half eigenenergies, $k \omega_f$ is the frequency of the $k^{\mathrm{th}}$ mode ($k\in \mathds Z$), $u_k$ are mode functions and $g$ is a coupling constant.
Solutions are found by dressing the spin progressively by each field mode. We begin by introducing a non-degenerate basis for the polychromatic field, then give a two frequency example, followed by the general case for an arbitrary number of frequency modes. The progressive dressing procedure is halted after $N-1$ transformations to give accurate analytic expressions for the time evolution operator of the system, even when the fields are strong compared to the frequency separation. Finally, a simple general expression for the time evolution of the spin is found for sufficiently weak fields.
Fields are considered with each mode initially in a coherent state, $\ket{\alpha_{k_i}}$, but the formalism is readily generalised for arbitrary initial states \cite{Glauber1963, Sudarshan1963}.
The Fock basis is energy degenerate when mode frequencies have a common multiple, $\omega_f$.
To avoid problems which arise from this degeneracy we decompose $\ket{ \{ \alpha_k \} }$ in a set of non-degenerate basis states \cite{Yuen2018}, 
\begin{equation} \label{eq:fieldstateNDB}
	\ket{\{ \alpha_k \}} = \sum_N \gamma_N \ket{N}.
\end{equation}
The basis state $\ket{N}$ is the normalised projection of $\ket{\{\alpha_k \}}$ onto subspace of Fock states $\ket{n_{k_1},n_{k_2},...}$ with energy $N \omega_f=\sum_k n_k k$. These eigenenergies form the same ladder of levels as the conventional approach but without including degenerate eigenstates.
Elsewhere it is shown that $\abs{\gamma_N}^2$ asymptotically approach a Gaussian distribution with mean $\sum_k k \abs{\alpha_k}^2$ and standard deviation $\sum_k k^2\abs{\alpha_k}^2$ \cite{Yuen2018}.

In this basis the quantum number operator is conveniently $\hat N = \sum_k k a_k^{\dagger} a_k$, and the creation and annihilation operator acts as $a^{\dagger}_k\ket{N}=(\gamma_N/\gamma_{N+k}) \alpha_k^* \ket{N+k}$ and $a_k\ket{N}=(\gamma_N/\gamma_{N-k}) \alpha_k \ket{N-k}$. For simplicity the mean field approximation $\gamma_N/\gamma_{N\pm k}\approx1$ is made. This is valid provided $k \ll \sigma_N$, although the general formalism can be applied without this approximation.
%
%
In the non-degenerate basis the Hamiltonian is now
\begin{equation} \label{eq:undressedH}
	H = \hat N \omega_f + \tfrac12 \omega_0 \sigma_z 
		+ \frac12 \sum_k \left( \Omega_k b_{k} \sigma_+ + \Omega_k^* b_{k}^{\dagger} \sigma_- \right)
\end{equation}
where $\Omega_k = \sqrt2 g u_k \alpha_k / \sqrt2$, with commuting field operators defined by $b_k \ket{N}=\ket{N-k}$ and $b_k^{\dagger} \ket{N}=\ket{N+k}$. Equation \ref{eq:undressedH} describes the mean field polychromatic Rabi model for a set of positive and negative frequency modes symmetrically distributed around zero \eqnref{eq:undressedH}. For all positive frequencies \eqnref{eq:undressedH} is the polychromatic Jaynes Cummings model.


This letter presents general analytic expressions for the $N$ frequency dynamics when the spin-half particle is closest to resonance with highest frequency mode. The modes are labelled in ascending order by index $k=1,2,... \ ... N$, and their frequencies are $\omega_k=(j+m_k)\omega_f$ with $j,k \in \mathds Z$. The positive integers $m_k$ are arranged in ascending order such that the lowest frequency is $j \omega_f$ ($m_1=0$), and the highest frequency is $(j+m_N)\omega_f$.

It is shown that the Hamiltonian can be partially diagonalised through a sequence of $N-1$ transformations, dressing the states one mode at a time. The resultant interaction Hamiltonian exhibits a strong, near resonant interaction between the `N-1 times dressed states', and several weaker, off resonant interactions which can be ignored. The unitary evolution of the dressed states under the strong interaction is calculated, then the inverse transformation applied to find the time evolution in the original frame of \eqnref{eq:undressedH}.


\begin{figure}[t]
	\begin{center}
		\input{figure1.tex}
	\end{center}
\caption{ \label{fig:twomode}
Excitation probabilities as a function of $\tau$ in a two mode field. Solid black lines show excitation probability calculated by diagonalising \eqnref{eq:undressedH} numerically. Dashed red lines show probabilities using \eqnref{eq:twomodeU0}. Figure (a) shows $\abs{\bra{\tfrac12} U^{(0)}(\tau) \ket{-\tfrac12}}^2$ after taking the partial trace over the field. Figure (b), (c) and (d) show excitation from $\ket{N,-\tfrac12}$ to $\ket{N-j,\tfrac12}$, $\ket{N-(j+m),\tfrac12}$ and $\ket{N-(j-m),\tfrac12}$ respectively.
}
\end{figure}

To begin an example with $N=2$ is considered, working in dimensionless frequency units $\omega_f^{-1}$ and dimensionless time $\tau = \omega_f t$.
Moving to an interaction picture $\ket{\psi}\to \tilde{\ket{\psi}} = U_1 \ket{\psi}$ using $U_1 = \exp -i \left(\hat N + \tfrac12 j \sigma_z\right)\tau$, the interaction Hamiltonian is
\begin{equation} \label{eq:V0}
	V^{(0)} = \tfrac12 \Delta_1 \sigma_z 
		+ \frac12 \sum_{k=1} \left( \Omega_k e^{-i m_k \tau} b_{j+m_k} \sigma_+ + \mathrm{h.c.} \right),\\
\end{equation}
where $\Delta_1 = \omega_0-j\omega_f$.
In general the diagonalisation of terms $\tfrac12 \Delta_k \sigma_z + \frac12 \left( \chi_k b_{j+m_k} \sigma_+ + \chi_k^* b_{j+m_k}^{\dagger} \sigma_-\right) \rightarrow \tfrac12 \widetilde \chi \sigma_z$ is performed with the unitary operator \begin{equation} \label{eq:Sk}
	S_k = 
		\frac{(\Delta_k + \widetilde \chi_k) \mathds 1 - \chi_k  b_j \sigma_+ + \chi_k^* b_{j+m_k}^{\dagger} \sigma_-}
		{\left[2 \widetilde \chi_k (\Delta_k+ \widetilde \chi_k)\right]^{\frac12}},
\end{equation}
where $\widetilde \chi_k = \sqrt{\Delta_k + \abs{\chi_k}^2}$.
Thus, $V^{(0)}$ is dressed by the first mode by applying $S_1^{\dagger} V^{(0)} S_1$, with $\chi_1 = \Omega_1$.
The time dependence of the mode $k=2$ terms is then removed by moving to a new interaction picture with $U_2=\exp(-\tfrac12 i m_2 \sigma_z \tau)$.This yields the `once dressed' interaction Hamiltonian, $V^{(1)}=U_2^{\dagger} \left( S_1^{\dagger} V^{(0)} S_1 - \tfrac12 m_2 \sigma_z \right) U_2$, and the states dressed by mode $1$ evolve under $i \partial_\tau \ket{\psi}=V^{(1)} \ket{\psi}$. Explicitly,
\begin{IEEEeqnarray*}{rCl} \label{eq:V1}
	V^{(1)} &=& 
		\tfrac12 \Delta_2 \sigma_z + 
		\frac12 \left[  \Omega_2 \left( \tfrac12 \ubar{\chi}_1 e^{-i m_2 \tau }b_{m_2} \sigma_z \right. \right.  \\
	&& \left. \left.
		\ubar{\Sigma}_1 b_{j+m_2} \sigma_+ + \ubar{\delta}_1 e^{-2 i m_2 \tau} b_{j-m_2}^{\dagger} \sigma_- \right) 
		+\mathrm{h.c.}\right] \IEEEeqnarraynumspace \IEEEyesnumber
\end{IEEEeqnarray*}
where $\Delta_2 = \widetilde \chi_1 - m_2$, $\Sigma_1 = \tfrac12(\Delta_1+\widetilde \chi_1)$ and $\delta_1 = \tfrac12(\Delta_1 - \widetilde \chi_1)$. Under bars represent values divided by $\widetilde \chi_k$, e.g. $\ubar{\chi}_1 = \chi_1 / \widetilde \chi_1$. 

In this `once dressed' picture the second mode is resonant when $\Delta_2=0$. 
Sufficiently close to this resonance, \eqnref{eq:V1} can be approximated by
\begin{equation} \label{eq:V1approx}
	V^{(1)}=
		\tfrac12 \Delta_2 \sigma_z + \frac12 \left( \chi_2 b_{j+m_2} \sigma_+ +  \chi_2^* b_{j+m_2}^{\dagger} \sigma_- \right),
\end{equation}
where $\chi_2 = \Omega_2 \ubar {\Sigma}_1$, since $\ubar \Sigma_1$ approaches unity when $\Omega_1<\omega_f$ and $\Delta_1\gtrsim0$.  In contrast, $\ubar \Omega_1$ and $\ubar \delta_1$ approach zero as $\Delta_2$ increases from -$\omega_f$. In principle, further dressing transformations could be made to include the effect of the weak terms ignored here.

The time evolution operator in the once dressed picture under \eqnref{eq:V1approx} is 
\begin{IEEEeqnarray*}{rCl}
	U^{(1)} &=& 
		\mathds 1 \cos \tfrac12 \widetilde \Omega_2 \tau \\
	&&
		- i \left( \ubar{\Delta}_2 \sigma_z + \ubar{\chi}_2 b_{j+m_2} \sigma_+ + \ubar{\chi}_2^* b_{j+m_2}^{\dagger} \sigma_- \right)
		\sin \tfrac12 \widetilde \chi_2 \tau .
\end{IEEEeqnarray*}
Transformed by $U^{(0)} = S_1 U_2 U^{(1)} S_1^{\dagger}$ to the undressed picture of \eqnref{eq:V0} yields
\begin{IEEEeqnarray*}{rCl}
	U^{(0)}(\tau) &=& 
		\mathds 1 \left( \cos \tfrac12 m_2 \tau \cos \tfrac12 \widetilde \chi_2 \tau - 
			\ubar{\Delta}_2 \sin \tfrac12 m_2 \tau \sin \tfrac12 \widetilde \chi_2 \tau \right) \\
	&&
		- i \sigma_z \left[
			\ubar \Delta_1 f(\tau)
			-\tfrac12 \sin \tfrac12 \widetilde \chi_2 \tau  
			\hat f_z(\tau)
		\right] \\
	&&
		- i \sigma_+ \left[
			\ubar \chi_1 b_j f(\tau)+
			\hat f_+(\tau)
			\sin \tfrac12 \widetilde \chi_2 \tau
		\right] \\
	&&
		- i \sigma_- \left[
			\ubar \chi_1^* b_j^{\dagger} f(\tau)+
			\hat f_-(\tau)
			\sin \tfrac12 \widetilde \chi_2 \tau
		\right] \IEEEyesnumber \label{eq:twomodeU0all} \IEEEyessubnumber \label{eq:twomodeU0}
\end{IEEEeqnarray*}
with $\hat f_-=f_+^{\dagger}$ and
\begin{IEEEeqnarray*}{rCl}
	f&=&\sin \tfrac12 m_2 \tau \cos \tfrac12 \widetilde \chi_2 \tau 
		+ \ubar{\Delta}_2 \cos \tfrac12 m_2 \tau \sin \tfrac12 \widetilde \chi_2 \tau 
		\IEEEyessubnumber \IEEEeqnarraynumspace \\
	\hat f_z&=&
		\ubar \chi_1^* \ubar \chi_2 e^{-\frac12 i m_2 \tau} b_{m_2}
			+ \ubar \chi_1 \ubar \chi_2^* e^{\frac12 i m_2 \tau} b^{\dagger}_{m_2} \IEEEyessubnumber \\
	\hat f_+&=&
		\ubar \Sigma_1 \ubar \chi_2 e^{-\frac12 i m_2 \tau} b_{j+m_2}
		+ \ubar \delta_1 \ubar \chi^*_2 e^{\frac12 i m_2 \tau} b_{j-m_2}. \IEEEyessubnumber
\end{IEEEeqnarray*}

The time evolution can be followed using the density operator $\boldsymbol \rho(\tau) = U^{(0)}(\tau) \boldsymbol \rho_i {U^{(0)}}^{\dagger} (\tau)$, where $\boldsymbol \rho_i = \ket{\psi_i} \bra{\psi_i}$. $\ket{\psi_i}$ is the initial state of the system at $\tau=0$, taken here to be $\ket{\{\alpha_k\},-\tfrac12}$. 
The spin-half density operator is found by taking the partial trace over the field, $\boldsymbol \rho_{\frac12}=\sum_N\bra{N} \boldsymbol \rho (\tau) \ket{N}$, and the excitation probability is
\begin{equation} \label{eq:densityoporator}
\bra{\tfrac12}\boldsymbol \rho_{1/2} \ket{\tfrac12}= 
 	\sum_N \abs{\sum_{N'} \gamma_{N'} \bra{N,\tfrac12}U^{(0)}(\tau)\ket{N',-\tfrac12}}^2.
\end{equation}
In \eqnref{eq:densityoporator} we can set $\gamma_{N'}=\gamma_{N''}$ since $U^{(0)}$ dictates that $N'$ and $N''$ differ from $N$ by $\pm m_1,\ \pm j$ or $\pm (j\pm m_1)$, and these differences are small compared to $\sigma_N$.
Hence the excitation probability is
\begin{equation} \label{eq:Pe}
	P_e(\tau) = \abs{\sum_{N'} \bra{N', \tfrac12} U^{(0)}(\tau) \ket{N,-\tfrac12}}^2.
\end{equation}
Substituting \eqnref{eq:twomodeU0all} into \eqnref{eq:Pe},
\begin{equation}
	P_e(\tau) = \abs{\ubar \chi_1 f(\tau) + f_+(\tau) \sin \tfrac12 \widetilde \chi_2 \tau)}^2
\end{equation}
where $f_+(\tau) = \sum_{N,N'} \bra{N'} \hat f_+(\tau) \ket{N}$.

Figure \ref{fig:twomode}a shows $P_e(\tau)$ from \eqnref{eq:twomodeU0} and \eqnref{eq:Pe} is in close agreement with numerical calculaions for the two frequencies $j \omega_f$ and $(j+2)\omega_f$ with $\Omega_1 = \Omega_2=\omega_f/2$ and $\omega_0 = (j+1)\omega_f$.
$P_{e}(\tau)$ is dominated by a sinusoidal oscillation at frequency $\widetilde \chi_2$, with smaller yet significant oscillations due to off resonant excitations. 
Figure \ref{fig:twomode}b,c\&d show the excitation probability between basis states $\abs{\bra{N',\tfrac12}U^{(0)}\ket{N,-\tfrac12}}^2$ which contribute to $P_{e}$. 
Figure \ref{fig:twomode}(b) shows the main contribution is between the resonant states with $N'=N-(j+2)$. The least accurate analytic term is between the off resonant state where $N'=N+j-2$, shown in \figref{fig:twomode}d. However, these terms give the smallest contribution with magnitude $\lesssim 5 \times 10^{-4}$.


These results depend on $m_2$, but not $j$. The choices $m_2=2$ and $j=-1$ represent two fields at frequencies $\pm \omega_f$, which is equivalent to a single mode interaction beyond the rotating wave approximation \cite{Rabi1937}, since $b_{-n}=b^{\dagger}_n$ in the mean field limit. Thus, the results \eqnref{eq:twomodeU0all} and \eqnref{eq:Pe} and their generalisations given below also produce accurate solutions to \eqnref{eq:undressedH} beyond the rotating wave approximation if positive and negative frequency modes are included symmetrically.

\begin{figure}[t]
	\begin{center}
		\input{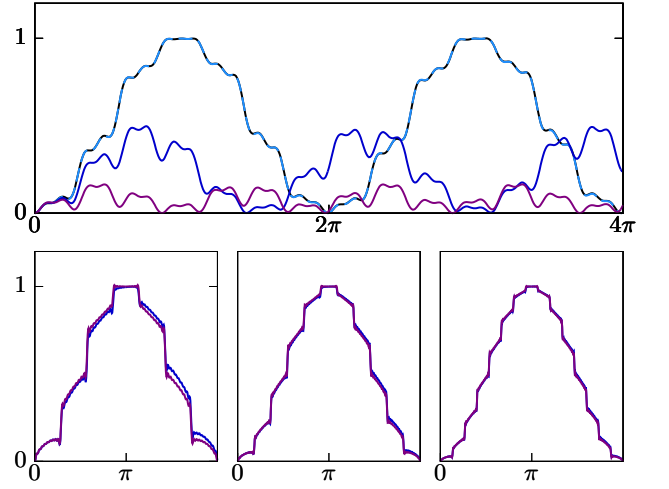}
	\end{center}
\caption{\label{fig:threeandtenmode}
Excitation probabilities in three mode (a) and ten mode (b-d) fields.
(a) $P_e(\tau)$ for three modes with $m_1=1,m_2=2$ and $\Omega_1=\Omega_2=\Omega_3=\omega_f/7$ for three detunings $\Delta_0=2\omega_f$ (light blue dashed), where the spin-half is resonant with mode $j+2$, $\Delta_0=\tfrac{13}{7}\omega_f$ (dark blue dashed) and $\Delta_0=\tfrac{6}{7}\omega_f$ (purple dashed). The analytic (coloured lines) and numerical (black lines) results are indistinguishable in these plots.
}
\end{figure}

This approach is now generalised to $N>2$ frequencies.
The interaction between the $k$ times dressed states is
\begin{equation} \label{eq:Vtransformation}
	V^{(k)} =
	U_{k+1}^{\dagger} \left(S^{\dagger}_{k} V^{(k-1)} S_{k} - \tfrac12 \delta m_{k+1} \sigma_z \right) U_{k+1}
\end{equation}
where $S_k$ is given by \eqnref{eq:Sk} and $U_k  = \exp\left(-\tfrac12 i \theta_k \sigma_z \right)$
with $\delta m_k=m_k - m_{k-1}$ and $\theta_k = \delta m_k \tau$.
Starting from $V^{(0)}$, $V^{(1)}$ can be found. 
From $V^{(1)}$ one finds the detuning $\Delta_2 = \widetilde \chi_1 - m_2$, and the coefficient of $b_{j+m_2} \sigma_+$ as $\chi_2= \ubar \Sigma_1 \Omega_{m_2}$. From these, $\widetilde \chi_2 = \sqrt{\Delta_2^2 + \abs{\chi_2}^2}$. 
$S_1$ is now given explicitly and $V^{(2)}$ subsequently found by applying \eqnref{eq:Vtransformation}. This process is repeated $N-1$ times to reveal the interaction resonant with the $N^{\mathrm{th}}$ mode between the $N-1$ times dressed states.

The transformation $U_{k+1}^{\dagger} S^{\dagger}_{k} V^{(k-1)} S_{k} U_{k+1}$ only changes the spin operators in $V^{(k-1)}$;
\begin{IEEEeqnarray*}{rCl}
	\sigma_z &\rightarrow& 
		\ubar \Delta_k \sigma_z - \ubar \chi_k \Theta_{k+1} b_{j+m_k} \sigma_+
		- \ubar \chi^*_{k} \Theta_{k+1}^* b_{j+m_k}^{\dagger} \sigma_-
		\IEEEyesnumber \label{eq:sigmatransformations} \IEEEeqnarraynumspace \IEEEyessubnumber  \\
	\sigma_+&\rightarrow&
		\frac {\ubar \chi^*_{k}}{2} b^{\dagger}_{j+m_k} \sigma_z + \ubar \Sigma_{k} \Theta_{k+1} \sigma_+
		+ \ubar \delta^*_{k} \Theta_{k+1}^* b_{2(j+m_k)}^{\dagger} \sigma_- \ 
		\IEEEeqnarraynumspace \IEEEyessubnumber
\end{IEEEeqnarray*}
The transformations, \eqnref{eq:Vtransformation}, is simple to apply successively when the interaction Hamiltonian is written as the vector $\mathbf v^{(0)} = \tfrac12(\Delta_0, \sum_m\Omega_m b_{j+m}, \sum_m \Omega_m^* b_{j+m}^{\dagger})^{\mathrm T}$ with basis $\boldsymbol \sigma = (\sigma_z, \sigma_+, \sigma_-)$. The vector is transformed by  \eqnref{eq:Vtransformation} through the vector equation
\begin{equation}
	\mathbf v^{(k)} = \mathbf M(k) \mathbf v^{(k-1)} - \tfrac12 \delta m_{k+1} \mathbf z
\end{equation}
where $\mathbf z = (1,0,0)^{\mathrm T}$. The matrix $\mathbf{M}(k)$ is
\begin{equation} \label{eq:Mkmatrix}
	\left(
		\begin{array}{ccc}
			\ubar \Delta_{k} & \tfrac12 \ubar{\chi}^*_{k} b^{\dagger}_{j+m_k} 
				& \tfrac12 \ubar \chi_{k} b_{j+m_k} \\
			-\ubar{\chi}_{k} \Theta_{k+1} b_{j+m_k} & \ubar \Sigma_{k} \Theta_{k+1}
				& \ubar{\delta}_{k} \Theta_{k+1} b_{2(j+m_k)} \\
			-\ubar{\chi}^*_{k} \Theta_{k+1}^* b^{\dagger}_{j+m_k} 
				& \ubar{\delta}^*_{k} \Theta_{k+1}^* b^{\dagger}_{2(j+m_k)} 
				& \ubar \Sigma_{k} \Theta_{k+1}^*
		\end{array}
	\right)
\end{equation}
where the columns are given by the coefficients of the terms on the right in \eqnref{eq:sigmatransformations}.
After each stage we can calculate $\Delta_{k+1} = \widetilde \chi_k - \delta m_{k+1} $ and
\begin{equation} \label{eq:chiN}
	\chi_{k+1} = \bra{n-(j+m_{k+1})} M_{2j}(k) v^{(k-1)}_j \ket{n}
\end{equation}
where lower indices label matrix and vector components.
Subsequently $\mathbf M(k+1)$ is specified using $\widetilde \chi_{k+1} = \sqrt{\Delta_{k+1}^2 + \abs{\chi_{k+1}}^2}$, $\Sigma_{k+1} = \tfrac12 (\Delta_{k+1} + \widetilde \chi_{k+1})$ and $\delta_{k+1} = \tfrac12(\Delta_{k+1} - \widetilde \chi_{k+1})$.

The final interaction Hamiltonian, dressed successively on modes $m_1, m_2,... \ ...,m_{N-1}$, can be approximated by the two level system
\begin{equation}
	V^{(N-1)} = \tfrac12 \Delta_N + \frac12 \left(\chi_N b_{j+m_N} \sigma_+ + \chi_N^* b_{j+m_N}^{\dagger} \sigma_- \right)
\end{equation}
where the truncated terms are off-resonant and of second order in $\Omega_j/\omega_f$ or higher.
The $N-1$ times dressed states evolve under
\begin{IEEEeqnarray*}{rCl}
	U^{(N)} &=& 
		\mathds 1 \cos \tfrac12 \widetilde \chi_N \tau - i \sin \tfrac12 \widetilde \chi_N \tau 
		\IEEEyesnumber \label{eq:UtauN}\\
	&&
		\times \left( \ubar{\Delta}_N \sigma_z + \ubar{\chi}_N b_{j+m_N} \sigma_+ 
		+ \ubar{\chi}_N^* b_{j+m_N}^{\dagger} \sigma_- \right)
\end{IEEEeqnarray*}
which is transformed back to the undressed picture by
\begin{equation}
	U^{(0)} = \left( S_1 U_{2}  \right)... \quad ...\left( S_{N-1} U_{N}  \right) U^{(N)} S_{N-1}^{\dagger}... \quad ... S_{1}^{\dagger}.
\end{equation}
Writing $U^{(N)}$ as the scalar product between the vectors $\boldsymbol \zeta = (\mathds 1, \sigma_z, \sigma_+,\sigma_-)$ and
\begin{equation}
	\mathbf u^{(N)} = \left(
	\begin{array}{c} 
		\cos\tfrac12 \widetilde \chi_N \tau \\
		-i \ubar \Delta_N \sin \tfrac12 \widetilde \chi_N \tau \\
		-i \ubar \chi_N b_{j+m_N} \sin \tfrac12 \widetilde \chi_N \tau \\
		-i \ubar \chi_N^* b_{j+m_N}^* \sin \tfrac12 \widetilde \chi_N \tau 
	\end{array} \right)
\end{equation}
this inverse transformation is given by
\begin{equation} \label{eq:u0vector}
	\mathbf u^{(0)} = \left[\prod_{k=1}^{N-1} \mathbf T(k) \right] \mathbf u^{(N)}.
\end{equation}
with transformation matrix
\begin{widetext}
\begin{equation} 
	\mathbf T (k) = \left(
	\begin{array}{cccc}
		\cos \frac12 \theta_{k+1} & - i \sin \frac12 \theta_{k+1} & 0 & 0 \\
		-i \ubar \Delta_k \sin \frac12 \theta_{k+1} & \ubar \Delta_k \cos \frac12 \theta_{k+1} 
			&-\tfrac12 \ubar \chi_k^*  e^{-\frac12 i \theta_{k+1}} b_{j+m_k}^{\dagger}
			&-\tfrac12 \ubar \chi_k e^{ \frac12 i \theta_{k+1}} b_{j+m_k} \\
		-i \ubar \chi_k \sin \frac12 \theta_{k+1} b_{j+m_k}
			&\ubar \chi_k \cos \frac12 \theta_{k+1} b_{j+m_k}
			&\ubar \Sigma_k e^{-\frac12 i \theta_{k+1}} 
			& \ubar \delta_k e^{\frac12 i \theta_{k+1}} b_{2(j+m_k)}\\
		-i \ubar \chi_k^* \sin \frac12 \theta_{k+1} b^{\dagger}_{j+m_k}
			&\ubar \chi_k^* \cos \frac12 \theta_{k+1} b^{\dagger}_{j+m_k}
			& \ubar \delta^*_k e^{-\frac12 i \theta_{k+1}} b_{2(j+m_k)}^{\dagger}
			&\ubar \Sigma_k e^{\frac12 i \theta_{k+1}}
	\end{array} \right)
\end{equation}

\end{widetext}

For $N$ frequencies the excitation probability, \eqnref{eq:Pe}, is the square modulus of the element $u_3^{(0)}$ of \eqnref{eq:u0vector} with field operators $b_k$ and $b_k^{\dagger}$ set to unity by the partial trace in \eqnref{eq:densityoporator}.
Figure \ref{fig:threeandtenmode}(a) shows extremely good agreement of $P_e(\tau)$ with numerical calculations for a three mode field for three different detunings. When the spin-half is resonant with the highest frequency field the oscillations at frequency $\widetilde \chi_3$ dominate $P_e(\tau)$. The smaller, higher frequency oscillations on top of these are driven by the off-resonant modes.
As $\omega_0$ decreases the $\widetilde \chi_3$ component reduces and oscillations driven by the second field mode begin to dominate.

When the Rabi frequencies are small compared to the mode spacing, $\mathbf M(k)$, $\mathbf T(k)$ and subsequent expressions can be simplified significantly by neglecting terms which contribute to $U^{(0)}$ beyond second order in $\Omega_k/\omega_f$. As above, a spin-half closest to resonance with the $N^{\mathrm{th}}$ mode is considered. One can set $\Delta_k = \Delta_0 - m_k \forall k$ and $\tilde \chi_k = \Delta_k$ for $k<N-1$ without affecting $U^{(0)}$ to second order in $\Omega_k/\omega_f$. Consequently $\Sigma_k=1$ and $\delta_k=0$. Iterating \eqnref{eq:chiN},
\begin{equation}
	\chi_N = \Omega_{m_N} \prod_{k=0}^{N-1} \ubar \Sigma_k + \mathcal{O}\left[(\Omega_j/\omega_f)^3\right]
\end{equation}
which simplifies to $\chi_k = \Omega_k$ with the approximations above. The time evolution operator in the $N-1$ times dressed basis is given by \eqref{eq:UtauN} with $\Delta_N=\Delta_0-m_N$, $\chi_N=\Omega_N$ and $\widetilde \chi_N = \sqrt{\Delta_N+\abs{\chi_N}^2}$  - one cannot approximate $\tilde \chi_N$ by $\Delta_N$ since it cannot be assumed $\Delta_N \gg \chi_N$. 
The multi-frequency effects retained must arise from the inverse transformations $\mathbf T(k)$. These are expanded to first order in $\Omega_k/\omega_f$ as
\begin{equation} \label{eq:Texapnsion}
	\mathbf T(k) = 
		\tfrac12 e^{\frac{i}{2} \theta_{k+1}} \left[ \mathbf T_0^+ + \mathbf T_k^+\right] 
		+ \tfrac12 e^{-\tfrac{i}{2} \theta_{k+1}} \left[ \mathbf T_0^- + \mathbf T_k^- \right],
\end{equation}
\begin{IEEEeqnarray*}{cc}
	\mathbf T_0^+ = \left( \begin{array}{cccc}
			1&-1&0&0 \\
			-1&1&0&0 \\
			0&0&0&0 \\
			0&0&0&2
		\end{array} \right), &
	\mathbf T_k^+ = \left( \begin{array}{cccc}
			0&0&0&0\\
			0&0&0&-\ubar \chi_k \\
			-\ubar \chi_k&\ubar \chi_k&0&0 \\
			-\ubar \chi_k^*&\ubar \chi_k^*&0&0
		\end{array} \right) \\
	\mathbf T_0^- = \left( \begin{array}{cccc}
			1&1&0&0 \\
			1&1&0&0 \\
			0&0&2&0 \\
			0&0&0&0
		\end{array} \right), &
	\mathbf T_k^- = \left( \begin{array}{cccc}
			0&0&0&0\\
			0&0&-\ubar \chi_k^*&0 \\
			\ubar \chi_k&\ubar \chi_k&0&0 \\
			\ubar \chi_k^*&\ubar \chi_k^*&0&0
		\end{array} \right),
\end{IEEEeqnarray*}
after taking the partial trace over the field operators.
The spin-half time evolution operator, accurate to second order in $\Omega_k/\omega_f$, is given by inserting \eqnref{eq:Texapnsion} into \eqnref{eq:u0vector}, and keeping terms to first order in $\mathbf T_k^{\pm}$. This gives
\begin{IEEEeqnarray*}{rCl} \label{eq:U0approx}
	U_{ge}^{(0)}(\tau) &=& \IEEEyesnumber \IEEEyessubnumber
		-i \ubar \chi_N e^{-i(N-1)\theta} \sin \tfrac12 \widetilde \chi_{N} \tau \quad \\
	&&
			+ \sum_{p=1}^{N-1} \frac{\ubar \chi_p}{2} \left[
			e^{i(N-2p-2)\theta} f_-(\tau)	
			+ e^{-i (N-1)\theta} f_+(\tau) \right]
\end{IEEEeqnarray*}
where $U_{ge}^{(0)}(\tau) = \bra{\tfrac12}U^{(0)}(\tau) \ket{-\tfrac12}$, and
\begin{IEEEeqnarray*}{rCl} \IEEEyessubnumber
	f_\pm(\tau) &=& 
		\pm \cos \tfrac12 \widetilde \chi_{N} \tau - i  \ubar \Delta_{N} \sin \tfrac12 \widetilde \chi_{N} \tau.
\end{IEEEeqnarray*}

Figures \ref{fig:threeandtenmode}(b)-(d) show $P_e(\tau)$ calculated analytically using \eqnref{eq:U0approx} with purples lines, and numerically with blue lines. The field has ten modes of equal amplitude, which corresponds to a pulse train the time domain. The mode amplitudes are $\omega_f/7$ in \figref{fig:threeandtenmode}(b), $\omega_f/11$ in (c) and $\omega_f/15$ in (d). This field drives an oscillations in the $P_e(\tau)$ with period $\widetilde 2\pi \chi_N^-1$. The oscillations are divided into either 7, 11 or 15 sloped plateaus separated by steep steps. The three mode resonant excitation in \figref{fig:threeandtenmode}(a) already shows this structure emerging with plateaus not yet smoothed out by farther off resonant fields. It is sensible to conclude that $P_e(\tau)$ approaches this stepped structure as the number of modes increases.
For ten modes, figures (b)-(d) show that the first and last plateaus of each cycle maintain a degree of curvature. The middle plateau is flat and very close to unity when $\omega_f/\Omega$ is odd. The occurrence of the steps is coincident with the pulses in the driving field. Interestingly, the sloped of the plateaus demonstrates that the excitation probability continues to evolve in-between pulses where the driving field is close to zero. Figures \ref{fig:threeandtenmode}(b)-(d) show the accuracy of \eqnref{eq:U0approx} increases as the mode amplitudes decrease, which is to be expected for an expansion in powers of $\Omega_k/\omega_f$.

In summary, accurate analytic expressions for the time evolution operator of a spin-half in a polychromatic second quantised field have been derived. A polychromatic dressed state formalism is presented to derive these results by progressively dressing states on each field mode. Furthermore, a simple closed form expression is derived for the spin-half's time evolution in a field with an arbitrary number modes which is valid when the mode amplitudes are sufficiently small compared to the mode spacing.

This work was supported by EU H2020 Collaborative project QuProCS (Grant Agreement No. 641277).

\bibliography{MultiFreqDressedAtoms}

\end{document}